\documentclass[sigconf]{acmart}

\usepackage{algorithmic}
\usepackage{comment}
\usepackage{graphicx}
\usepackage{textcomp}
\usepackage{xcolor}
\usepackage{bchart}
\usepackage{tabularx}
\usepackage{graphicx}
\usepackage{hyperref}
\hypersetup{
    colorlinks=true,
    linkcolor=blue,
    filecolor=blue,
    urlcolor=blue,
}

\usepackage[framemethod=TikZ]{mdframed}
\mdfdefinestyle{style1}{
innerleftmargin=0.2cm,innerrightmargin=0.2cm,
innertopmargin=0.2cm ,innerbottommargin=0.2cm,
roundcorner=10pt,linewidth=0.6pt,
footnoteinside=false}

\AtBeginDocument{%
  \providecommand\BibTeX{{%
    \normalfont B\kern-0.5em{\scshape i\kern-0.25em b}\kern-0.8em\TeX}}}

\acmConference[BotSE '20]{BotSE '20: International Workshop on Bots in Software Engineering}{May 24th, 2020}{Seoul, South Korea}
\begin{document}

%%
%% The "title" command has an optional parameter,
%% allowing the author to define a "short title" to be used in page headers.
\title{Challenges and guidelines on designing test cases for test bots}

%%
%% The "author" command and its associated commands are used to define
%% the authors and their affiliations.
%% Of note is the shared affiliation of the first two authors, and the
%% "authornote" and "authornotemark" commands
%% used to denote shared contribution to the research.
\author{Linda Erlenhov, Francisco Gomes de Oliveira Neto, Martin Chukaleski, Samer Daknache}
%\orcid{1234-5678-9012}
\affiliation{%
  \institution{Chalmers and the University of Gothenburg}
  \city{Gothenburg}
  \country{Sweden}
}
\email{linda.erlenhov@chalmers.se, francisco.gomes@cse.gu.se, {guschuma,gusdaksa}@student.gu.se}

%%
%% By default, the full list of authors will be used in the page
%% headers. Often, this list is too long, and will overlap
%% other information printed in the page headers. This command allows
%% the author to define a more concise list
%% of authors' names for this purpose.
\renewcommand{\shortauthors}{Erlenhov et al.}

%%
%% The abstract is a short summary of the work to be presented in the
%% article.
\begin{abstract}
Test bots are automated testing tools that autonomously and periodically run a set of test cases that check whether the system under test meets the requirements set forth by the customer. The automation decreases the amount of time a development team spends on testing. As development projects become larger, it is important to focus on improving the test bots by designing more effective test cases because otherwise time and usage costs can increase greatly and misleading conclusions from test results might be drawn, such as false positives in the test execution. However, literature currently lacks insights on how test case design affects the effectiveness of test bots. This paper uses a case study approach to investigate those effects by identifying challenges in designing tests for test bots. Our results include guidelines for test design schema for such bots that support practitioners in overcoming the challenges mentioned by participants during our study.
\end{abstract}

%%
%% The code below is generated by the tool at http://dl.acm.org/ccs.cfm.
%% Please copy and paste the code instead of the example below.
%%


%%
%% Keywords. The author(s) should pick words that accurately describe
%% the work being presented. Separate the keywords with commas.

%% A "teaser" image appears between the author and affiliation
%% information and the body of the document, and typically spans the
%% page.

%%
%% This command processes the author and affiliation and title
%% information and builds the first part of the formatted document.
\maketitle

\section{Introduction}

Testing is an essential activity performed throughout software development and maintenance. However, increasing complexity of software-intensive systems, in addition to resource constraints, hinder test effectiveness. A combination of software automation tools and bots help decrease the time that development teams spend on software testing and support testers\slash developers to make smarter decisions related to testing activities~\cite{storey_zagalsky_2016}. Literature refers to such tools as \textit{test bots}, and have been applied to leverage test coverage, test flakiness, and test planning~\cite{storey_zagalsky_2016}. Test bots are part of a wider range of software bots, particularly DevBots, which can be seen as artificial software developers that are autonomous, adaptive, and has technical competence~\cite{linda-devbots}. 

However, current literature lacks studies focusing on the challenges related to designing and applying test bots, such as the hindrances or utilities of test bots when designing test cases or planning test executions. Therefore, our goal is to investigate the current industry practices and challenges with software testing aided by test bots, particularly, test design practices. We perform a case study with an automotive company in Sweden where we (i) interview industry practitioners and (ii) analyse test artefacts to, respectively, identify their current practices and associated challenges, as well as to propose guidelines on how to design tests for their test bots. In short, our research questions are:

\begin{itemize}
   \item[RQ1:] What are the main challenges when designing and executing system tests on software bots?
   \item[RQ2:] To what extent does the test design affect the effectiveness of the test bot? 
\end{itemize}

Our contributions are: (i) a list of challenges reported by industry practitioners related to designing test cases for a test bot, and (ii) a guidelines with six items to support practitioners in designing test cases when using test bots. Particularly, factors such as execution time, cyclomatic complexity of test code and usage of synchronous\slash asynchronous programming affect the cost and maintenance of test bot in software testing. However, our findings are limited to the context of our case study, i.e., test bots used in system testing.

\section{Related work}
According to Lebeuf et al.\@ \cite{lebeuf_storey_zagalsky_2018} software bots can help improve the efficiency of every phase of the software development life cycle, including test coding. The paper outlines the different types of bots and how they can respectively help improve software development. While the paper is beneficial for outlining the difference between bots, it does not dive deeper into the different instances of bots, but rather provides an overview of how these bots can be beneficial. In turn, Erlenhov et al.\@'s \cite{linda-devbots} study proposes a taxonomy focused on DevBots while also providing definition and vision of future DevBots. Both taxonomies are relevant within our study as they provide insight on specific properties that helps distinguishing test bots from other software bot applications (e.g., chat bots). For instance, the test bots in our study fall in the group of productivity bots, because they improve the development team’s productivity by automating the execution of testing tasks~\cite{lebeuf_storey_zagalsky_2018} and interact with users via notifications on dashboards and team communication channels~\cite{linda-devbots}, whereas other facets such as language~\cite{linda-devbots} are not central at the current stage of our investigation.

Moreover, our contributions align with existing literature in test design practices, but targeting the specific aspects of test bots such as automation and autonomy. For instance, Tsai et al.\ \cite{tsai2001end} gives an overview of how to design tests, and depicts the creation of test scenario specification, test case generation and tool support. Authors illustrate the effort or time spent when creating end-to-end (E2E) system tests and the portion which is taken by the integration tests. Both types of test cases target verification of distinct parts and properties of a System Under Test (SUT) which, ultimately, impacts the test costs. Similarly, Mockus et al.~\cite{mockus_nagappan_dinh-trong_2009} investigate how test coverage affects test effectiveness and the relationship between test effort and the level of test coverage, whereas Laventhal et al.\@ \cite{leventhal1994analyses} discuss in their paper the relevance of negative and positive tests and how testers show positive test bias, which can affect the quality of testing. %Note that the common theme between our research and the aforementioned studies is the examination of test effectiveness along with test coverage, however, we aim to align\slash contrast those test design practices with the usage of test bots in industry.
Our results differ from already investigated problems in automated testing because our guidelines focus to enhance the test bot's autonomy and interaction, two facets that are distinguishing bots from plain test automation, where related work on test automation mainly targets the test's cost-effectiveness. Although both aspects are related, we discuss properties that accentuate the test bot's capabilities of acting as artificial developers.

%A case study was performed in a company within the software engineering industry using an investigative approach where practitioners were interviewed and software artefacts related to test bots were analyzed. The data collected from this study along with the analysis, was cross-evaluated and gave us information that guided us into creating a test design schema for test bots. The motivation behind why this research was performed is due to almost 50 to 70\% of effort from the total software development is spent on testing and approximately 50 to 70\% of it lies in the group of functionality testing \cite{tsai2001end}. Establishing a good test design schema for the automated test bots can potentially help provide better test coverage, which in turn can help increase test effectiveness \cite{mockus_nagappan_dinh-trong_2009}. The usage of the term test effectiveness implies that the test case should be simple, faster when executing and lastly less complex (e.g., in terms of cyclomatic complexity and lines of code). It is important to mention that this study does not aim to pursue \textit{defect detection rate}.
\section{Research methodology}
The summary of our case study planning is depicted in Table \ref{tab:planning}. Our case company is a relatively mature company that provides Services as a Product (SaaP) for different car manufacturing companies. Our research is going to be performed only within the scope of one project responsible for developing scalable software solutions for a specific car manufacturer.

\begin{table}
\centering
\caption{Case study planning according to guidelines by Runeson et al.~\cite{runeson2009guidelines}}
\label{tab:planning}
\vspace{-0.4cm}
\begin{tabularx}{\columnwidth}{lX}
 \toprule
 \textbf{Objective} & \textbf{Description}  \\ 
 \midrule
  The context & Black box, end-to-end system testing \\ 
  The case & One project from the automotive industry  \\ 
  Research Questions & RQ1 and RQ2 \\ 
  Theory  & Test case design, DevBots \\ 
  Methods & Direct and independent data collection \\
  Selection strategy & Project using test bots for system testing \\
  Analysis & Thematic analysis of interviews \\  
           & Qualitative assessment of artefacts \\  	
 \bottomrule
\end{tabularx}
\vspace{-0.5cm}
\end{table}

The software under test uses Amazon Web Services (AWS) and Microsoft Azure as their cloud service providers. The developed software is deployed in a virtual private cloud (VPC)~\cite{serrano2015Infrastructure} in order to offer customer companies the benefits of a private cloud, such as a granular control over virtual networks and an isolated environment for sensitive workloads and service isolation. The organization develops their own in house system which is responsible for building the software artefacts that are constantly updated by the developers. It supports multiple languages like Java, Python, GO, among others.

We performed a semi-structure interview with selected participants following standard protocols for data collection where participants were asked for consent to use their data and all the collected data was anonymised. Participants were also given the opportunity to opt-out of the study at any time. The interviewees were selected based on having previous experience with the test bots, which means that they were familiar with the overall scope of software test bots and had worked on their development. Four participants agreed to join our study, namely: one software architect, two senior and one junior software developer.

Our list of questions is available at: \url{https://tinyurl.com/botse2020}. The interviews were recorded, transcribed (upon consent from participants) and coded. We performed thematic analysis~\cite{braun_clarke_2006} on the interview data in order to find patterns in the raw data later used as the base for the coding \cite{maguire2017doing}. The outcome were categories which summarized the data gathered and expressed key themes and processes related to their usage of the test bots. Lastly, our findings were later presented to the participants of the interview to clarify and validate our understanding of their process.

Additionally, we also collected data from software artefacts which included test bot code, test case code and requirements for the SUT in order to investigate the design of the test bots and their test cases. Moreover, those two data sources (interviews and development artefacts) offer insights that enable us to answer, respectively, RQ1 and RQ2. The next section comprise our findings and discussion based on the interviews done with practitioners and the analysis of the artefacts related to the test bot. Based on our collected data, we begin by explaining how the test bots are used at our case company, followed by answer to our research questions and validity threats to our study.
\section{Findings and Discussion}

\begin{figure}
    \includegraphics[width=\columnwidth]{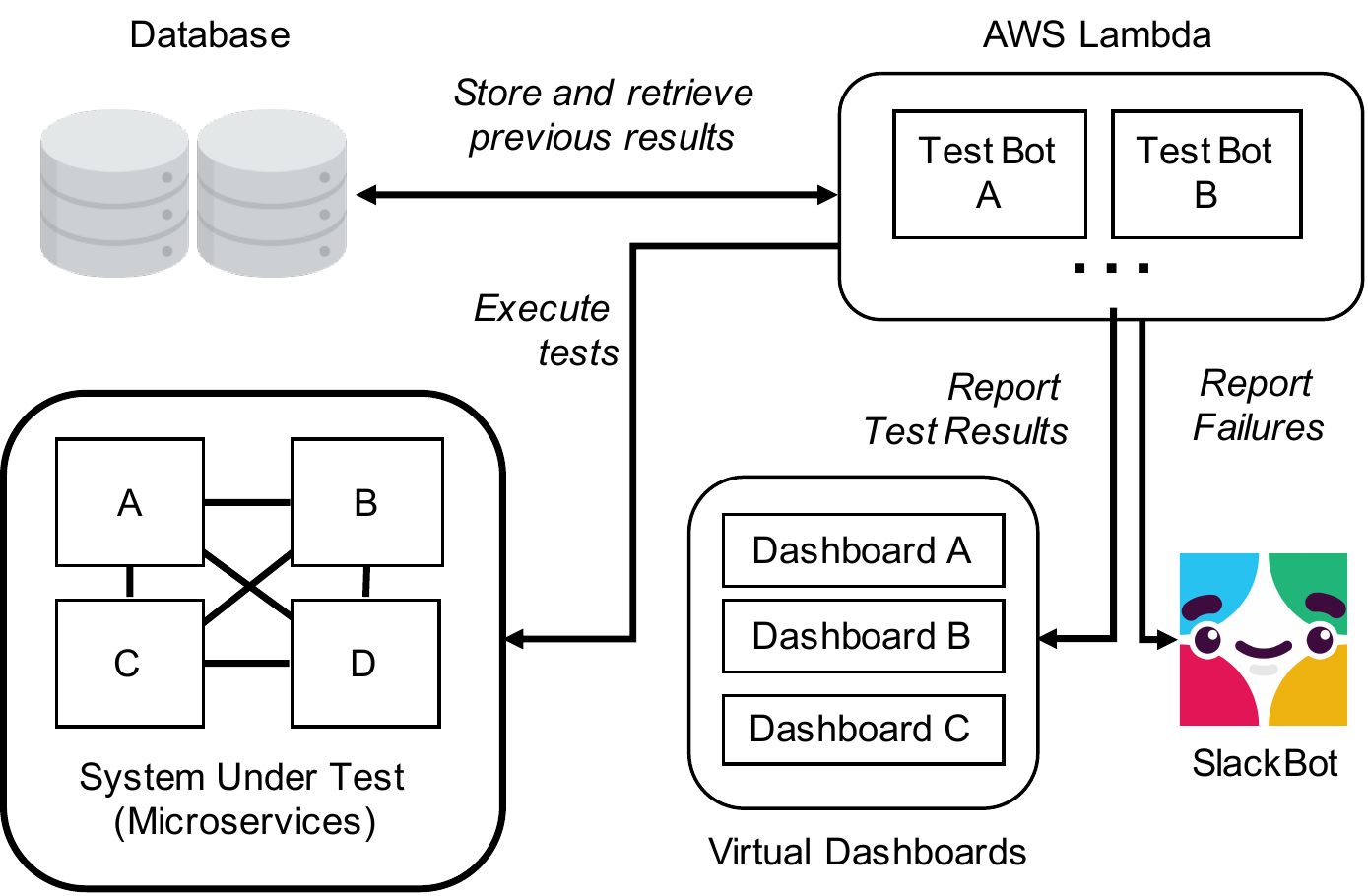}
    \centering\caption{Overview of the application of test bots.}
    \label{fig:testbot}
    \vspace{-0.8cm}
\end{figure}

The company creates various test bots for load testing, integration testing and system testing, however, for the scope of this study, we are going to focus only on test bots that are performing system testing. Figure \ref{fig:testbot} shows an overview of the test bots and the systems and tools it interacts with. The test bots used in the program have the task of performing end-to-end tests with a specific rate on different functionalities of the system. Depending on the testing context, whether the system that needs to be tested is back-end or front-end oriented, different programming languages will be used to write the test cases and the test bot. %For front-end services developers use JavaScript, while on the other hand for back-end services Java, Python and Scala are used. 

Consequently, in order to design the test cases, practitioners are required to have knowledge on the corresponding programming languages and an overview of the architecture of the component, in order to be able to understand the flow between the different microservices. The regular workflow of the bot involves the autonomously triggered execution of test cases, logging the test results and, in case of failures, include detailed error logs to make debugging easier. Lastly, the test bot interacts with the practitioners in two ways. First by submitting the results data to virtual dashboards shown to the entire team indicating the status of test activities. Secondly, if the test fails, the test bot uses slack to notify the corresponding team about the status and details of test execution.

%The responsibility of the test bots within the case study company goes beyond test execution, hence they are not responsible for creating mocked or simulated environments before running the tests, rather the bot is deployed within the environment that it needs to test. 

%One test bot can contain multiple test cases which test different services. For instance, consider testing the functionality of creating a new user in an online system. The several steps required to test this functionality, e.g., create a user with specific data and doing database checks for system state before and after addition, are all 

\begin{figure}
    \centering
    \includegraphics[width=\columnwidth]{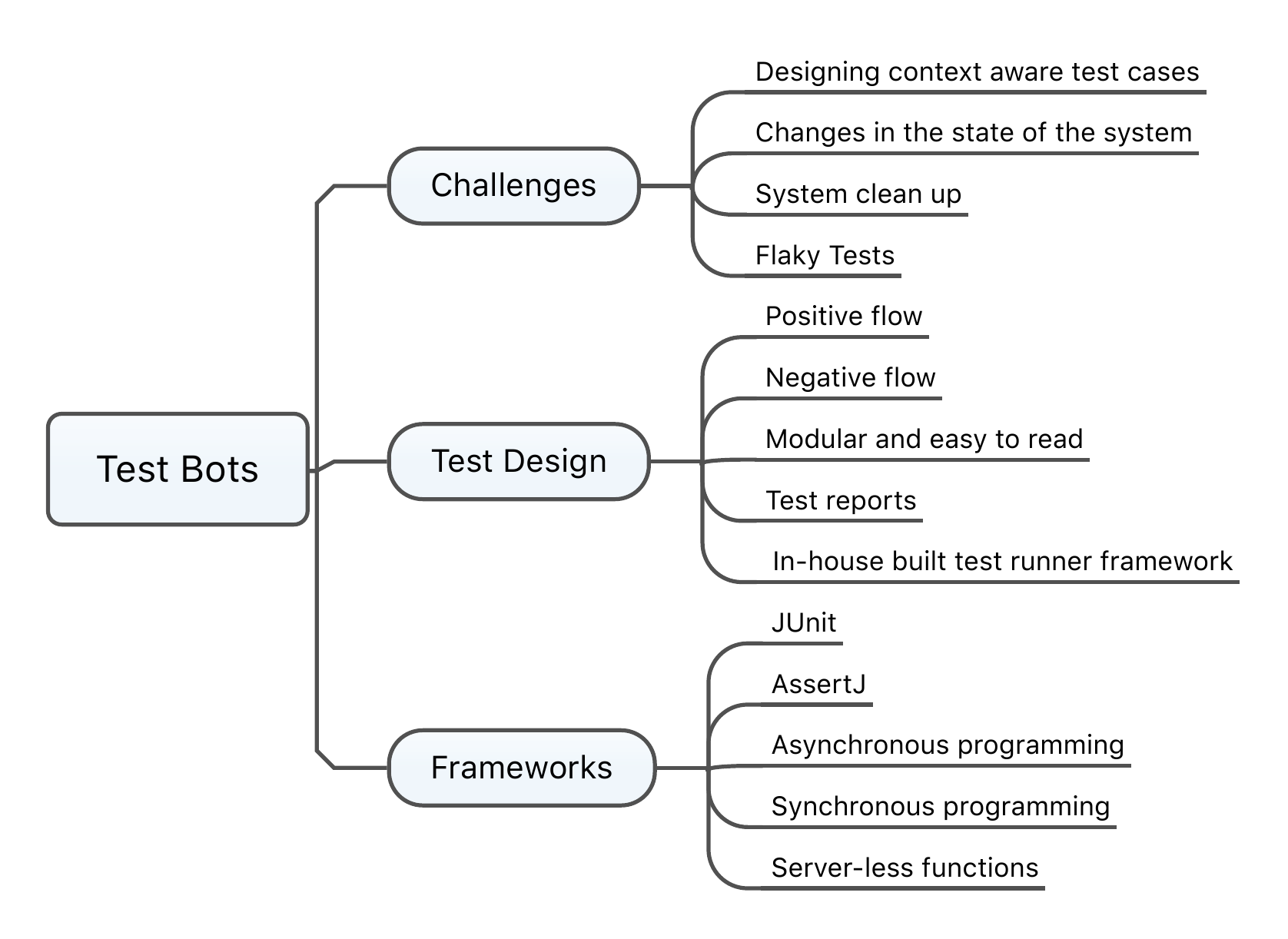}
    \vspace{-0.8cm}
    \caption{Themes and codes from the interview data.}
    \label{fig:testthemes}
    \vspace{-0.5cm}
\end{figure}

\subsection{Analysis of RQ1}

Figure \ref{fig:testthemes} shows the resulting themes and codes from our thematic analysis of the interviews. The data revealed different aspects about the test bots, pertaining their composition (e.g., test frameworks), usage (e.g., design of test cases for the test bots) and, in connection to our research question, the main challenges in applying them into the company's software testing activities.

\textit{C1. Designing context aware test cases}: 
Tests can have dependencies to other tests meaning that a particular test $t_a$ can only start executing once another test $t_b$ has finished successfully. Performing this specific chaining on test cases can become quite complex and timely to achieve using typical test frameworks, such as JUnit. Issues can arise in the event that a test case fails, thus the following dependent tests will be affected by the previous success rate, hence being hard to determine whether the test failed because of some dependency or faults in the code. This dependency should be avoided when designing the test cases, such that modular tests are preferred to yield more independent test executions.

\textit{C2. Changes in the state of the system}:
Issues can occur when a test suite has only partially executed leaving, then, corrupt or invalid data within the system, such as incomplete data models which can later cause system errors or null pointer exceptions. To mitigate this problem, the test bot needs to perform roll back techniques in order to remove the invalid data that has been generated.

\textit{C3. System clean up}: In scenarios when all of the test cases have completed successfully, the test bot needs to clean up after themselves, since the test bots run on the deployed production environment. Consequently, there is a risk to mix test bot activity with the customer activity, hence confusing developers while monitoring the application logs to debug a problem. For instance, in a scenario where the test bot is flooding the system with the test associated logs, it becomes difficult to find, among the logs, other issues that could be user related. While challenge C2 addresses recovering from corrupt data of individual tests, the clean up for this challenge would equate to a tear-down of all tests. Automated test frameworks often support creation of tear down methods where the system state is restored after executing the test suite.

\par \textit{C4. Flaky tests}: Flaky tests are false positives, i.e., test cases that fail when, in reality, there are no faults and the functionality is working correctly. Consequently, flaky tests consume a lot of the tester's time and effort~\cite{luo2014empirical}. At the case company, the SUT requires real time connections with different vehicles in order to collect their status, and tests often fail due to a lacking connection as opposed to a fault itself. Test containing asynchronous wait for connection is among the top categories of flaky tests~\cite{luo2014empirical}. Therefore, the stability of third party incorporated systems must be considered. One way to mitigate this problem is to mock the vehicle behaviour on a separate cloud server instance with very high uptime. Conversely, this can be costly and time-consuming to develop.\\

\begin{mdframed}[style = style1]
\textbf{RQ1:} In short, the challenges identified for system-level test bots are: (i) designing context aware test cases, (ii) monitoring and controlling changes in the state of the system, (iii) instrumenting system rollbacks and (iv) detecting flaky results.
\end{mdframed}

\begin{table*}[!ht]
    \caption{Guidelines for creating good system test case design}
    \vspace{-0.3cm}
    \label{table:guidlines}
    \centering
    \footnotesize
    \begin{tabularx}
    {\linewidth}
    {>{\hsize=.07\hsize}X %Column 1; 2.5% of 4
     >{\hsize=.8\hsize} X %Column 2; 20% of 4
     >{\hsize=1.6\hsize}X %Column 3; 40% of 4
     >{\hsize=1.53\hsize}X} %Column 4; 37.5% of 4
       % sum=4.0\hsize for 4 columns. Adjust the weights to sum
       % 4, but increasing/decreasing per column
    \toprule
    \textbf{ID}& \textbf{Description}& \textbf{Reason behind guideline}& \textbf{Suggestion on how to implement it} \\
    \midrule
    %G1  & Ensure completeness of system specifications and or documentation. & Testing a functionality of a system might result in positive results, in accordance to the function's purpose, however it might not take into consideration the system specification, and thus developers might miss an important testing aspect. & Include trace links between test cases and user documentation (e.g., include a Jira item ID into the test case). \\ \midrule
    G1 & Use asynchronous programming methods to invoke system endpoints. & According to the data analyzed, asynchronous programming methods allows the tests to continue testing independent functions simultaneously and can thus reduce execution time. & With the Java framework CompletableFuture test can be executed with the method. \texttt{supplyAsync()}, in this case the framework will run the task asynchronously and return the result from the test without blocking the execution of other test.\\
    \midrule
    G2 & Cover test dependencies by chaining dependant test via the specific callback asynchronous functions. & For dependant tests, tests which require the completion status of previous tests, this would allow the system to wait before executing the next tests, while for independent tests, these can be unchained thus allowing all the non dependant test to run within their order. & With the Java framework CompletableFuture test $A$ can be executed with the method \texttt{supplyAsync()}, and test $B$ needs to be chained to the first future using the function  \texttt{thenComposeAsync()}, this way test $B$ will execute once $A$ has finished without blocking other test executions in the test suite. \\
    \midrule
    G3 & Create test that are small, modular and readable. & By creating smaller and more readable tests, developers can ensure that future alterations to the tests are easier to implement, as well as less time will be needed to analyze the existing setup and develop the new test. & Create test cases that test a specific functionality of the system rather than multiple flows. \\
    \midrule
    G4 & Start clean-up process after the execution of tests. & In order to avoid adding corrupt or unnecessary test data to real environments, developers should include a clean-up process after the execution of tests within the test bot.  & Use existing (or implement) functionality to remove data which was stored in the system by the test suite (e.g., after testing the user creation functionality, use the delete user service to remove the test data). \\
    \midrule
    G5 & Make use of proper logging techniques which differentiate the test data from real data. &  In order to make it easier to distinguish the \textit{test activity logs} from the actual \textit{user activity logs}. & Use different prefix for test data attributes, thus making it easier for developers to distinguish test data from real data in real environments (e.g., username starts with ``TEST'').  \\
    \midrule
    G6 & Implement both positive and negative flow testing techniques. & According to Laventhal et al.\@ \cite{leventhal1994analyses}, software testing theory suggests that tests should test inside and outside specification (expected versus unexpected values) in order to test thoroughly. & Use both invalid and valid test data as input. \\
    \bottomrule
    \end{tabularx}
\end{table*}

\subsection{Analysis of RQ2}

Based on the artefacts analysis, we identified two different approaches to implement the test bots using, respectively, synchronous and asynchronous programming. Even though the different designs tested identical functionalities of the system, usage of asynchronous programming was more beneficial due to: (i) faster execution time, (ii) fewer false positive results and (iii) reduced cyclomatic complexity of the test code.

Regarding execution time, asynchronous allowed different tests to run in parallel (when applicable) leading to faster test execution. Consequently, test bots become cheaper for our industry partner due to the pay per use model of server-less applications where the test bots are hosted. Moreover, asynchronous approaches motivated testers remove dependencies between tests, hence mitigating issues with flaky tests. This factor improves the following DevBot aspects of test bots: (i) autonomy, since individual test bots can act independently from each others, and (ii) user interaction, since notifications and results sent to developers become more credible. Finally, the artefact analysis also revealed that less complex tests (regardless of being synchronous\slash asynchronous) were easier to maintain by practitioners making it easier to also maintain the test bot. Although asynchronous programming yields better test cases for a test bot, such approaches are coupled with the instrumentation offered by testing frameworks. In order to overcome this limitation, practitioners created a simple test runner class which would handle\slash monitor all the test cases, together with the creation of the test reports.\\

\begin{mdframed}[style = style1]
\textbf{RQ2:} Design choices affect execution time, flakiness and complexity of test cases executed by the test bot. Particularly, using asynchronous programming benefits the bot since it enables (i) parallel and faster execution of tests, and (ii) diligence in designing context-aware and independent test cases.
\end{mdframed}

\vspace{-0.3cm}
\subsection{Guidelines for Designing Tests for Test Bots}

Using the data collected from interviews and artefacts, we created guidelines (Table~\ref{table:guidlines}) that should mitigate the challenges discussed in the previous section. The guidelines foster good system test cases for test bots similar to the ones used at our industry partner. We also provide a reason for including each guideline along with suggestions on how to implement it.

Additionally, our findings also relate to existing literature on guidelines to design test cases, such as creating reusable and readable tests (G3)~\cite{tsai2001end} or targeting both positive and negative flows for tests (G6)~\cite{leventhal1994analyses}. In contrast, literature does not consider unique aspects of test bots, such as autonomy or interaction with users, when discussing test design practices to an extent where one can argue whether current design practices are relevant to a DevBot.

The discussion around challenges also triggered participants to share what kind of future improvements they would like to add to the test bot functionality. An example is to predict the time needed for the system to perform each functionality in the test suite. Later, this information can be combined with the history of test results to reveal patterns able to describe how usage spikes can affect system performance. This data is valuable and can be used to extensively configure the application for improved performance.

\subsection{Threats to Validity}

In turn, \textit{construct validity} is associated to our choice of artefacts, participants, themes and corresponding codes. One of the main risks is that literature lacks consolidated constructs or definition of a bot hindering the proper selection of bots for a study~\cite{linda-devbots}. We mitigate this threat by using existing taxonomies~\cite{linda-devbots,lebeuf_storey_zagalsky_2018} to identify whether the investigated test bots have properties based on the facets of those taxonomies.

Our \textit{internal validity} is related to the interview process and the analysis of artefacts, such as overlooking relevant aspects of participant's answers. In order to mitigate risks, the interviews were recorded and transcribed. Moreover, we used a semi-structured format such that participants were allowed to slightly stray from the questions in order to convey their own understanding of the process that the questions may fail to capture. Moreover, the review of artefacts and interpretation of the test bot process described in the interviews was later validated with our industry partners. 

In turn, our \textit{external validity} is limited, since our conclusions and findings are connected to the case company's context. Future studies investigating similar aspects of test bots applied to domains beyond automotive industry can confirm\slash contrast our findings. Moreover, clearer definitions of a bot and their role in software development can support generalization in future studies and enable researchers to find the commonalities between the applicability of bots across different domains of software development.
\vspace{-0.3cm}
\section{Conclusion}

%Our interview with industry practitioners and analysis of software artefacts revealed a set of challenges faced when designing test cases for system tests executed by a test bot, such as: designing context aware tests, monitoring and controlling system states, performing system clean-ups and identifying flaky tests. Moreover, we provide a list of guidelines to support practitioners in designing test cases for test bots similar to ones used in our context where bots are hosted in a cloud infrastructure communicating with vehicles. For instance, our results indicate that usage of asynchronous programming improves test effectiveness since (i) it reduces testing time and costs, and (ii) mitigates risks with flaky tests. Future studies include the investigation of other types of bots used by our industry partner, as well as investigating the impact of those design practices in other types of test bots focusing, e.g., integration and performance testing.

Our interview with industry practitioners and analysis of software artefacts revealed a set of challenges faced when designing test cases for system tests executed by a test bot, such as: designing context aware tests, monitoring and controlling system states, performing system clean-ups and identifying flaky tests. Moreover, we provide a list of guidelines to support practitioners in designing test cases for test bots similar to ones used in our context where bots are hosted in a cloud infrastructure communicating with vehicles. For instance, our results indicate that usage of asynchronous programming improves the effectiveness of a test bot since (i) it reduces time to execute tests and the overall testing costs, and (ii) brings awareness to issues with dependencies between tests, hence mitigating risks with flaky tests. Future studies include the investigation of other types of bots used by our industry partner, as well as investigating the impact of those design practices in other types of test bots focusing, e.g., integration and performance testing.

\bibliographystyle{IEEEtran}
\bibliography{botse2020_paper.bib}
\end{document}